\documentclass[a4paper,twocolumn]{article}
\usepackage{amsmath}
\usepackage{amsfonts}
\usepackage{amssymb}
\usepackage{graphicx}
\usepackage{natbib}
\usepackage{authblk}
\usepackage{color}

\begin{document}

\title{The heating history of Vesta and the onset of differentiation}

\author[1,2]{Michelangelo Formisano\thanks{e-mail address: michelangelo.formisano@iaps.inaf.it}}
\author[1,3]{Costanzo Federico}
\author[1]{Diego Turrini}
\author[1]{Angioletta Coradini}
\author[1]{Fabrizio Capaccioni}
\author[1]{Maria Cristina De Sanctis}
\author[3]{Cristina Pauselli}
\affil[1]{INAF-IAPS, Via del Fosso del Cavaliere 100, 00133 Rome (Italy)}
\affil[2]{University of Rome ``La Sapienza'', Piazzale Aldo Moro 5, 00185 Rome (Italy)}
\affil[3]{Dept. of Earth Science - University of Perugia, 06123 Perugia (Italy)}

\date{}

\maketitle

\begin{abstract}

In this work we study the link between the evolution of the internal structure of Vesta and thermal heating due to $^{26}$Al and $^{60}$Fe and long-lived radionuclides, taking into account the chemical differentiation of the body and the affinity of $^{26}$Al with silicates.

We explored several thermal and structural scenarios differing in the available strength of energy due to the radiogenic heating and in the post-sintering macroporosity. By comparing them with the data supplied by the HEDs and the Dawn NASA mission, we use our results to constrain the accretion and differentiation time as well as the physical properties of the core.

Differentiation takes place in all scenarios in which Vesta completes its accretion in less than 1.4 Ma after the injection of $^{26}$Al into the Solar Nebula. In all those scenarios where Vesta completes its formation in less than 1 Ma from the injection of $^{26}$Al, the degree of silicate melting reaches 100 vol.$\%$ throughout the whole asteroid. If Vesta completed its formation between 1 and 1.4 Ma after $^{26}$Al injection, the degree of silicate melting exceeds 50 vol.$\%$ over the whole asteroid but reaches 100 vol.$\%$ only in the hottest, outermost part of the mantle in all scenarios where the porosity is lower than 5 vol.$\%$. If the formation of Vesta occurred later than 1.5 Ma after the injection of $^{26}$Al, the degree of silicate melting is always lower than 50 vol.$\%$ and is limited only to a small region of the asteroid.

The radiation at the surface dominates the evolution of the crust which ranges in thickness from 8 to about 30 km after 5 Ma: a layer about 3-20 km thick is composed of primitive unmelted chondritic material while a layer of about 5-10 km is eucritic.
\end{abstract}


\section{Introduction}

Is Vesta a terrestrial world? And if so, of which kind? Vesta is the most geologically diverse of the large asteroids, showing a surface geology as differentiated as the one of the Moon and Mars as revealed by resolved images taken by the the Hubble Space Telescope (\Citealt{Gaffey,Binzel1,Li}) and recently confirmed by the spectroscopic measurements of the Dawn mission (\Citealt{DeSanctis3}).
Surface spectroscopy of Vesta indicates regions that are basaltic (see, e.g., \Citealt{McCord,Gaffey,Binzel1,Li,DeSanctis3}), therefore due to ancient lava flows and effusive phenomena. This is surprising evidence that Vesta once had a molten interior, like the Earth does. After the crust formation, Vesta underwent an intense and extended resurfacing due to collisional evolution (\Citealt{Turrini1,Coradini2}), as can also be inferred by the presence of a large crater on the south pole (\Citealt{Schenk}).

From a spectroscopic point of view, most of the terrains of Vesta have a typical pyroxene signature (\Citealt{McCord,DeSanctis3}), as revealed by the presence of the 0.9 and 1.9 $\mu$m absorption bands. The presence and the shapes of these bands in the spectra of Vesta, moreover, match those in the spectra of the Howardite, Eucrite, Diogenite meteorites (HEDs in the following, see, e.g., \Citealt{McCord,Gaffey,DeSanctis3}), a match that led to the consideration of Vesta as the source of the HEDs collected on the Earth (see \Citealt{McSween11} for a more detailed discussion).
This would make Vesta one of few intact bodies in the Solar System that differentiated and for which rock samples are available. Therefore Vesta can be used as a case study to investigate the primordial stages of the evolution of the terrestrial planets.

The differentiation of asteroids requires a strong heat source, capable of producing high temperatures in bodies with high surface-to-volume ratios. The heat source must have also operated quite early in the history of the Solar System so that the differentiation could have occurred within the first few millions years of the Solar System (see, e.g., \Citealt{Scott07}).

Several heat sources have been proposed: the only one that currently seems to be able to meet the requirements, is the heat produced by the decay of short-lived radioactive isotopes. Among the possible isotopes the most effective both in terms of concentration and of the production of energy seems to be $^{26}$Al (\Citealt{Urey}). \citet{Bizzarro} suggested that the accretion of chondritic parent bodies $\sim$1.5 Ma after the formation of Calcium-Aluminum-rich Inclusions (CAIs in the following) is the limit to initiate the melting.

A wide range of thermal models of planetesimals with $^{26}$Al as the heat source exists in the literature: we briefly analyze the most significant ones, describing first the models concerning Vesta and subsequently those regarding generic asteroids.

The first study on the differentiation of asteroid 4 Vesta with a numerical code was presented by \citet{GMC}. These authors demonstrated that it is possible to sustain partial melt on Vesta for $\sim$100 Ma assuming radiogenic heating by $^{26}$Al. They examined the influence of the delay time of accretion with respect to the formation time of the CAIs and they assumed instantaneous accretion. They showed that melting and core formation did not occur if the time of accretion is more than 3 Ma after CAIs formation, based on the assumption that the eucrites were generated with no more than 25 vol.$\%$ partial melting of silicate (\Citealt{Stolper}). They did not use a fixed temperature at the surface, where the heat loss is governed by the energy balance between the radiation at the surface and the interior heating due to the decay of short radionuclides. They did not examine the evolution of the internal structure such as the percolation of the iron through the silicatic matrix and consequent core formation.

\citet{Gupta} analyzed the differentiation path of Vesta and of other differentiated asteroids of sizes 20-270 km, assuming a linear accretion rate and introducing parameterized convection in the molten core and magma ocean. They included sintering but they used a constant temperature surface boundary. The authors explored two different scenarios dealing with the origin of the basaltic achondrites: in the first the origin is linked to partial silicate melting, in the second to the residual melt left after the crystallization in a cooling magma ocean. The core-mantle differentiation in both the scenarios initiated subsequent to 40 vol.$\%$ silicate melting.

In literature we also find many papers dedicated to the study of accretion and differentiation of generic asteroids.

\citet{Merk} examined the influence of the accretion process on the thermal evolution. They concluded that the accretion rate must be considered as long as the accretion time was not much smaller than the half-life of the radionuclide under study.

\citet{HS} also incorporated convection in their thermal evolution models of planetesimals assuming that mantle regions where the degree of partial melting exceeded 50 vol.$\%$, i.e. a magma ocean (\Citealt{Taylor2}), would be convecting. They simulate mantle convection by assuming that at temperatures greater than 1725 K the thermal conductivity increases by three orders of magnitude. As a result, heat is transferred rapidly from the convecting interior into the overlying rigid, partially molten zone. These authors modeled the primordial history of the planetesimals by analyzing the sintering process when the temperature reaches the value of $\sim$700 K. The authors concluded that parent bodies of differentiated meteorites accreted within 2 Ma from CAIs before most chondritic parent bodies and they suggested that molten planetesimals may be a source for chondrule melt droplets.

\citet{Sahijpal} performed numerical simulations of the differentiation of planetesimals undergoing a linear accretion growth with both $^{26}$Al and $^{60}$Fe as the heat sources. They studied in particular the dependence of the growth rate of the Fe, Ni-FeS core on the onset time of planetesimal accretion (relative to CAIs formation), the (constant) accretion rate, the final size of the planetesimal, and the $^{60}$Fe$/^{56}$Fe initial ratio. They did not use a radiation boundary condition but they included the sintering and linear accretion. The authors developed two scenarios: in the first, the core grew gradually before silicate began to melt and in the second the core segregated once the silicate had become 40$\%$ molten. In their model they showed that the rate and time of core and crust formation depend strongly on the onset time of the planetesimal accretion: significant melting occurred when accretion was completed before 2 Ma.

Recently, \citet{Mosko} studied how the migration of silicate melt and in particular the redistribution of $^{26}$Al from the interior into a crustal layer would affect the thermal evolution of planetesimal. In their model, core formation would require a bulk melting degree of 50 vol.$\%$. They only considered the case of instantaneous accretion but they did not include convection or sintering and they used a fixed temperature at the surface. The authors concluded that differentiation would be most likely for planetesimals larger than 20 km in diameter and that accreted within approximately 2.7 Ma from CAIs.

\citet{Sramek} presented a multiphase model for differentiation of planetesimals which takes into account phase separations by compaction driven melt migration. According to these authors, for the bodies with radii greater than 500 km, impacts provide an additional heat source. They included accretion, sintering, radiation boundary condition, analyzing the evolution of the internal structure. The authors concluded that, depending on the accretion rate, melting may start either in the center, or directly under the surface, or both: as noted by \citet{Merk}, accretion phase plays a crucial role in the thermal history of planetesimals.

\citet{Neumann} focused on the differentiation of small planetesimals ($<$ 120 km) for melt fractions smaller than 50 vol.$\%$. They combined the calculation of conduction, accretion, sintering, melting and melt segregation by porous flow. The authors showed that the differentiation of planetesimals depends principally on the accretion time and metal and silicate segregation are almost simultaneous processes and last between 0.4 and 10 Ma.

As in \citet{GMC}, \citet{Merk}, \citet{Mosko} and \citet{Sramek} we do not consider sintering and prefer to analyze the post-sintering phase, exploring different cases of porosities. This is due to the large uncertainties associated to the assumed initial porosities and the corresponding thermal conductivities. The initial value (0.001 W m$^{-1}$K$^{-1}$) assumed by \citet{HS} and \citet{Sahijpal} is based on laboratory measurements of the lunar regolith in a vacuum (\Citealt{Fountain}) and, as pointed by \citet{Mosko}, we do not consider lunar regolith as an ideal analog for an unsintered planetesimal.
Similarly we do not model accretion, as in \citet{GMC}, \citet{HS} and \citet{Mosko}. As pointed by \citet{Mosko}, recent dynamical studies treating the turbulent concentration of small particles in proto-planetary disks (\citealt{Johansen07,Johansen09}; \Citealt{Cuzzi}) showed that planetesimals can grow ``nearly instantaneously" in $<$ 100 yr to sizes of 100 km or larger. These timescales are much smaller than the half-life of the radionuclides here considered so the accretion time can be considered negligible. We also neglect the heat released by the accretion process since it is insignificant relative the short-lived radionuclide contribution (in the unrealistic case in which all the kinetic energy is converted in heat, the maximum temperature increase possible is about 90 K). As in \citet{Neumann} and \citet{Sramek} we use a Darcy's law model for segregation of metal and silicates, which is a distinct advance over the earlier Vesta-specific models of \citet{GMC} and \citet{Gupta}. The affinity of $^{26}$Al with silicates causes, in our model, a different distribution of the heat in the evolving structure: as a consequence temperatures reach the maximum value in the mantle enriched in silicates, as we observe in the thermal profiles of \citet{GMC} but not in \citet{Gupta}. We also assume a radiation boundary condition, unlike of \citet{Gupta}, which led to the study of the evolution of the crust and the depiction of a reliable heating history of Vesta. As \citet{Merk} we include Stefan law formulation to incorporate the effect of latent heat into the evolution.

In the current work we assume as a critical parameter the delay time $\Delta t_d$, which incorporates both the delay in the injection of $^{26}$Al in the Solar Nebula and the temporal interval covered by the accretion time of Vesta. The delay time $\Delta t_d$ is important because it determines the content of radioactive material at the onset of the thermal evolution of Vesta. The longer the delay time, the weaker is the available source.
We also develop sub-scenarios characterized by different post-sintering porosity in order to improve the treatment of the internal structure of the rocky asteroid, in particular the formation and the evolution of the core, by studying its physical and chemical properties, and to depict the primordial history of $4$ Vesta (and in general of all rocky asteroid partially or completely differentiated), constraining the accretion and differentiation times.
This work provides a theoretical support to data analysis for the Dawn mission, which addresses the investigation of the Vesta's internal structure as well as the composition of the crust and of the underlying mantle in the region of giant impact basins.

\section{Geology and Geochemistry}

Vesta formed at the same time and in the same Solar System environment as the terrestrial planets, but the asteroid never grew to planetary size because of the formation of Jupiter at $5.2$ AU, which caused disruptive resonances in the Main Belt. Recent papers investigated the bombardment on Vesta due to Jupiter's formation and to its migration (\Citealt{Turrini1,Turrini2}). Their results clearly indicated that the formation of the giant planet caused an intense early bombardment in the orbital region of the Main Belt. In the most extreme scenario, for Jupiter migrating by 1 AU, the basaltic crust of Vesta would not have survived intact to the Jovian Early Bombardment (\Citealt{Turrini2}). In the other studied cases, Vesta's surface is saturated by craters as big as 100 km and its crust would have significantly eroded (\Citealt{Turrini1,Turrini3}).

Evidence for rapid iron-silicate differentiation of asteroids, i.e. the formation of the core, comes from $^{182}$Hf-$^{182}$W concentration variations in iron meteorites (\Citealt{Horan,Kleine}).
\citet{Mark} confirmed that metal-silicate differentiation must have occurred on the parent body of magmatic irons no later than $\sim$1 Ma after CAIs.
Early differentiation is also supported by the presence of excesses of $^{26}$Mg from the decay of extinct $^{26}$Al in angrites (\Citealt{Bizzarro}) and
in the eucrites \emph{Piplia Kalan} (\Citealt{Sri}) and \emph{Asuka 881394} (\Citealt{Nyq}).

Using $^{182}$Hf-$^{182}$W data for eucrites, \citet{Kleine} showed that mantle differentiation in Vesta (assumed the parent body of eucrites) occurred at 4563.2$\pm$1.4 Ma and they suggested that core formation took place 0.9$\pm$0.3 Ma before mantle differentiation.

Moreover, the compositions of HEDs have been used to estimate the asteroid's bulk composition and core mass (\Citealt{McSween11,Zuber}). The evidence for the formation of a metallic core is provided by the siderophile elements depletions in HEDs relative to chondrites. \citet{Rig2} estimated the mass of the core in the range 5-25$\%$ of the total mass, based on the models of siderophile elements abundances in chondrites; otherwise, for \citet{Ruzicka} the mass of the core is in the range 4-30$\%$ of the total mass, based on mixtures of HED minerals that produce chondritic element ratios and using a bulk density of $\sim$3540 kgm$^{-3}$, slightly different with respect to the most recent value (3456 kgm$^{-3}$) given by \citet{Russell}. \citet{Ruzicka} estimated a metallic core of Vesta of $<$130 km in radius.

The olivine may be a significant component of the mantle, based on the occurrence of a limited number of olivine-bearing diogenites and a dunite having HEDs oxygen isotopic composition (\Citealt{McSween11}). \citet{Ruzicka} suggested an olivine-rich mantle $\thicksim$65-220 km thick.

The crust of Vesta consists of some mixture of basalts (eucrites) and a significant fraction of ultramafic cumulates (diogenite). \citet{Warren} estimated the mixing ratio of basaltic eucrite to diogenite in howardite regolith breccias to be constant at 2:1. The normative mineralogy of that composition is $\sim$70 vol.$\%$ of pyroxene and $\sim$25 vol.$\%$ of plagioclase, corresponding to a value for density of $\sim$3170 kgm$^{-3}$. This ratio, as suggested by \citet{Warren}, could be an upper limit (and the real crustal proportions may be closer to 1:1) because of bias in the impact sampling of deeper lithologies. \citet{Ruzicka} suggested a lower crustal unit ($\thicksim$12-43 km thick) composed of pyroxenite from which diogenites were derived and an upper crustal unit ($\thicksim$23-42 km thick) from which eucrites originated.

While we know that Vesta is a differentiated body, the exact mechanism for the formation of its core is still a matter of debate. In fact, in literature, there are two schools of thought regarding core separation from silicatic matrix. The first scenario for the formation of the core is that the melt fraction of the silicates is required to be larger than about 50 vol.$\%$ (\Citealt{Taylor1,Taylor2}), arguing for the presence of an early magma ocean in a planetesimal to form a core. This assumption is supported by experimental studies that partial melting of meteorites does not show metal migration (\Citealt{Takahashi,Walker}). The second scenario suggests that iron segregation and possibly core formation can start already for small melt fractions of iron (\Citealt{Larimer,Hewins}) even before silicate starts to melt. This assumption is supported by the observations of Fe, Ni-FeS veins in the acapulcoite-lodanite parent body (\Citealt{McCoy}) and by recent experiments suggesting an interconnected melt network for pressures below 2-3 GPa (\Citealt{Tera}).

\section{The Model}
\subsection{Initial and Boundary Conditions of the Model}

Our model is based on 1D heat conduction with radiogenic heat source and with black-body radiation at the surface. We assume Vesta as a spherical body of fixed radius equal to 270 km and initially composed of a homogenous mixture of two components, the first one ($\sim$25 vol.$\%$) generically referred to as \emph{metals} (essentially Fe and Fe-S) and the second ($\sim$75 vol.$\%$) generically referred to as \emph{silicates}. This composition is similar to those of the H and L classes of the ordinary chondrites, which contain significant amounts of metals (\Citealt{McSween0}) even if the inferred composition for Vesta is slightly different, as it appears to be strongly depleted in sodium and potassium (\Citealt{ConsolDrake}). The post-sintering porosity ranges from 1 to 5 vol.$\%$. We opt for these values of porosity because the lower bulk density estimate derived from Dawn suggests 5 to 6 vol.$\%$ in the mantle and crust, which is consistent with Vesta's intense bombardment history as revealed by Dawn (\Citealt{Russell}).

The initial temperature ($T_0$) of the body (that is also the local temperature of the Solar Nebula) is fixed to 200 K (\Citealt{Lewis}): a change of $T_0$ to 300 K does not affect the results in any significant way.
During the thermal evolution, the silicatic and metallic fractions per unit of volume change as a consequence of differentiation. The physical parameters density, specific heat and thermal conductivity of each volume unit also change accordingly. We impose a radiation boundary condition at the surface and a \emph{Neumann boundary condition} (heat flux equal to zero) at the center, as expressed by the following equations:

\begin{equation}\label{eq:T0}
 T(r,t = 0) = T_0,
\end{equation}

\begin{equation}\label{eq:partialT0}
\left[ {\dfrac{{\partial T}}{{\partial r}}} \right]_{r = 0}  = 0,
\end{equation}

\begin{equation}\label{eq:rbc}
\left[ \dfrac{\partial T}{\partial r}\right]_{surf} = - \dfrac{\varepsilon \sigma}{K} \left(T_{surf}^4 - T_0^4\right),
\end{equation}
where $T_{surf}$ is the temperature of the surface, $\varepsilon$ is the emissivity and $\sigma$ is the Stefan-Boltzmann constant (see Table \ref{tab:Parameters}).

\subsection{Physical Description of the Model}

To numerically study the thermal evolution of Vesta we take into account the heating due to decay of $^{26}$Al, $^{60}$Fe and long-lived radionuclides (e.g. $^{238}$U,$^{235}$U). The initial concentrations of the two short-lived radioactive elements $^{26}$Al and $^{60}$Fe together with their half-lives are reported in Table \ref{tab:Parameters}.

The equation of heat transfer in a porous medium, assuming local thermal equilibrium so that $T_{sol} = T_{liq} = T$ (here \emph{sol} stands for solid and \emph{liq} for fluid taking averages over an unit of volume), becomes (following \Citealt{Nield}):
\begin{equation}\label{eq:heat}
{\left( {\rho c} \right)_m}\dfrac{{\partial T}}{{\partial t}}  = \vec \nabla  \cdot \left( {{K_m}\vec \nabla T} \right) + H,
\end{equation}
where
\begin{equation}
\begin{array}{l}
 {\left( {\rho c} \right)_m} = \left( {1 - \phi } \right){\left( {\rho c} \right)_{sol}} + \phi {\left( {\rho c} \right)_{liq}} \\
  \\
 {K_m} = \left( {1 - \phi } \right){K_{sol}} + \phi {K_{liq}} \\
 \end{array}
\end{equation}
are the overall heat capacity and the overall thermal conductivity respectively; $H$ is the overall heat production per unit volume of the medium.
The surface temperature ($T_{surf}$) is controlled by the radiation boundary condition (see the eq.\eqref{eq:rbc}).
The volumetric radiogenic heating rate, due to $^{26}$Al decay, following \citet{Castillo1}, can be expressed as:
\begin{equation}\label{eq:rad}
{H_{Al}} = \bar{\rho} {C_{Si}}{\left[ {{}^{26}Al} \right]_0}{H^ * }{e^{ - \lambda t}},
\end{equation}
where $\bar{\rho}$ is the mean density, $C_{Si}$ is the mass fraction of the silicates, $[^{26}$Al]$_0$ is the initial concentration of $^{26}$Al in kg for kg of silicates, $\lambda =ln(2)/\tau_{Al}$ is the decay constant and $H^{*}$ is the specific power production (see \Citealt{Castillo1} and Table \ref{tab:Parameters}). An analogous formulation holds for $^{60}$Fe and long-lived radionuclides (for decay information see Table of \Citealt{Castillo1}).

Once the melting temperature of Fe-FeS (or of silicates) is reached, partial or complete melting occurs depending on the parameter
\begin{equation}
\chi  = \frac{{T - T_{sol}}}{{T_{liq} - T_{sol}}},
\end{equation}
following \citet{Merk} and assuming a linear growth of $\chi$ with raising temperature ($T$). The values of the temperature for the initial ($T_{sol}$) and complete ($T_{liq}$) melting temperature of metals (or of silicates) are reported in Table \ref{tab:Parameters}.
Also following \citet{Merk}, the specific heat (that we assume not to depend on the temperature) is modified through the Stefan coefficient,
\begin{equation}
Ste =  \dfrac{L}{c}\dfrac{{d\chi }}{{dT}} = \dfrac{L}{c}\dfrac{1}{{T_{liq} - T_{sol}}},
\end{equation}
to take into account in a simple way the latent heat during phase transition:
\begin{equation}
\bar{c} = c(1 + Ste).
\end{equation}
As in \citet{GMC}, iron melting is initiated at 1213 K, the melting temperature of the eutectic Fe-FeS system, and silicate melt generation is assumed to initiate at 1425 K (see Table \ref{tab:Parameters}). The entire latent heat for melting is assumed to be expended in a temperature ``window" between \emph{solidus} and \emph{liquidus} (\Citealt{GMC}). This simplification of temperature ``windows" does not change the whole thermal history by being the exact latent heat supplied to cause silicate and metallic melting.
The percolation of the metals through the silicate matrix, starting at 1213 K, is governed by the advection equation. If we assume $Y$ to be the concentration of the metals, the equation reads:
\begin{equation}\label{eq:chim}
\dfrac{{\partial Y}}{{\partial t}} + \vec v \cdot \vec \nabla Y = 0,
\end{equation}
in which $v$ is the migration velocity of the molten metal, given by Darcy's law. The chemical diffusion is assumed negligible.

The migration velocity of molten metal, following \citet{Yoshino1,Yoshino2} and \citet{Senshu2}, can be expressed:
\begin{equation}\label{eq:vel}
v = \dfrac{{{K_D}}}{\mu }g\Delta \rho
\end{equation}
where $K_D$ is the permeability of the silicate medium, $\mu$ = 0.005 Pa$\cdot$s is the viscosity of molten iron, $g$ is the gravitational acceleration, $\Delta \rho$ is the density contrast between molten metal and solid silicate (see Table \ref{tab:Parameters}).
The permeability-porosity relationship is expressed by:
\begin{equation}\label{eq:permeability}
K_D = \dfrac{\phi^n r_g^2}{\beta},
\end{equation}
where $\phi$ is the porosity, $r_g =$ 10$^{-3}$ m is the grain size, $\beta = 200$ is a geometrical constant and $n =$ 2 is predicted in an isotropic model with regular pore network along the edge of grains with tetrakaidecahedron form (\Citealt{Yoshino2}).

As noted by \citet{Yoshino1}, if permeable flow is established, segregation velocities (which vary in the range 1-100 m/a) are rapid in comparison with the timescale of core formation predicted from the $^{182}$Hf-$^{182}$W isotope system. During the percolation the volume fraction of the metallic component ($Y$) goes down enriching the forming core region. When the temperature reaches the value of the 50 vol.$\%$ melting temperature of the silicate (assumed to be 1725 K, \Citealt{Taylor1}), the separation of two melts occurs and the silicate component ($X$) moves upwards to the mantle region dragging $^{26}$Al with it. At the end of this phase, the core becomes pure metallic because the melted metals, being more dense than silicatic ones, sinks to the center.
During the evolution, the concentration of Al grows in the mantle underneath the lithosphere while the density profile varies due to the differentiation and the moment of inertia factor decreases starting from the initial characteristic value of 0.4 for a uniform sphere.

We investigate several evolutive scenarios, varying the strength of the radiogenic sources through the delay time ($\Delta_t$) and the porosity. The combined solution of the eqs.\eqref{eq:heat} and \eqref{eq:chim} led us to study the evolution of the internal structure, constraining formation time, size and mass of the core of Vesta, the size of the ``crust" and the temperature profile as a function of the distance from the center and of the time.
\begin{table*}[!htcp]
\resizebox{1\textwidth }{!}{
\begin{tabular}[c]{lccc}
  \hline
  \textbf{Quantity}  & \textbf{Value} & \textbf{Unit} & \textbf{Reference}\\
  \hline
  & & & \\
  Final primordial radius ($R$) & 270$\times10^{3}$ & $m$ & \Citealt{GMC}\\
  Density of metal (solid) ($\rho_{met,sol}$) & 6300 & $Kg$ $ m^{-3}$ & \Citealt{Neumann}\\
  Density of silicate (solid) ($\rho_{sil,sol}$) & 3200 & $Kg$ $ m^{-3}$ & \Citealt{Neumann}\\
  Density of metal (liquid) ($\rho_{met,liq}$) & 6200 & $Kg$ $ m^{-3}$ & \Citealt{Neumann}\\
  Density of silicate (liquid) ($\rho_{sil,liq}$) & 2900 & $Kg$ $ m^{-3}$ & \Citealt{Neumann}\\
  Specific heat of metal (solid) ($c_{met,sol}$) & 600 & $J Kg^{-1} K^{-1}$ & \Citealt{Sahijpal}\\
  Specific heat of metal (liquid) ($c_{met,liq}$) & 2000 & $J Kg^{-1} K^{-1}$ & \Citealt{Sahijpal}\\
  Specific heat of silicate(solid) ($c_{sil,sol}$) & 720 & $J Kg^{-1} K^{-1} $ & \Citealt{Sahijpal}\\
  Specific heat of silicate (liquid) ($c_{sil,liq}$) & 720 & $J Kg^{-1} K^{-1}$ & \Citealt{Sahijpal}\\
  Latent heat of metal ($L_{met}$) & 270 & $ KJ $  $ Kg^{-1}$ & \Citealt{GMC}\\
  Latent heat of silicate ($L_{sil}$) & 400 & $ KJ $  $ Kg^{-1}$& \Citealt{GMC}\\
  Metal solidus ($T_{sol}^{met})$ & 1213 & K & \Citealt{GMC}\\
  Metal liquidus ($T_{liq}^{met})$ & 1233 & K & \Citealt{GMC}\\
  Silicate solidus ($T_{sol}^{sil}$) & 1425 & K & \Citealt{Taylor1}\\
  Silicate liquidus ($T_{liq}^{sil}$) & 1850 & K & \Citealt{Taylor1}\\
  50 vol.$\%$ of silicate melting temperature
  ($T_{50}^{sil}$) & 1725 & K & \Citealt{HS}\\
  Thermal conductivity of metal ($K_{met}$) & 50 & $W$ $m^{-1}$ $K^{-1}$ & \Citealt{Sramek}\\
  Thermal conductivity of silicate ($K_{sil}$) & 3 & $W$ $m^{-1}$ $K^{-1}$ & \Citealt{Sramek}\\
  Initial metal volume fraction ($Y$)& $25\%$&  & \\
  Initial silicate volume fraction ($X$) & $75\%$& & \\
  Post-sintering porosity ($\phi$) & $1\%$ - $5\%$ &  & \\
  Temperature of Solar Nebula ($T_0$) & 200 & $K$ & \Citealt{Lewis} \\
  Stefan-Boltzmann constant ($\sigma$) & 5.67$\times 10^{-8}$ & $W$ $m^{-2}$ $K^{-4}$ &\\
  Emissivity ($\varepsilon$) & 1 &  &\\
  Half-life of $^{26}Al$ ($\tau_{Al}$) & $0.717$ & $Ma$ & \Citealt{Castillo2}\\
  Heat production of $^{26}Al$ & $0.355$ & $W Kg^{-1}$ & \Citealt{Castillo2}\\
  Initial isotopic abundance of $^{26}Al$ in ordinary chondrites ($[^{26}Al]_0$)& $6.20\times10^{-7}$ & $ppb$ & \Citealt{Castillo2}\\
  Half-life of $^{60}Fe$ ($\tau_{Fe}$) & $2.62$ & $Ma$ & \Citealt{Rugel}\\
  Heat production of $^{60}Fe$ & $0.068 - 0.074$ & $W Kg^{-1}$ & \Citealt{Castillo1}\\
  Initial isotopic abundance of $^{60}Fe$ in ordinary chondrites ($[^{60}Fe]_0$)& $(22.5 - 225) \times10^{-9}$ & $ppb$ & \Citealt{Castillo1}\\
  & & & \\
  \hline
\end{tabular}}
\caption{Physical parameter values used in this work.}\label{tab:Parameters}
\end{table*}

\subsection{Numerical Procedure}

The numerical solution of the system of differential equations \eqref{eq:heat} and \eqref{eq:chim} is obtained using a 1D finite difference method (Forward-Time Central-Space, \emph{FTCS}) in radial direction. A spatial grid of $\Delta r = 300$ m is used. To avoid numerical stability problems due to the instability of the FTCS scheme we use the \emph{Lax correction} (\Citealt{Num}).

To ensure the stability of our numerical approach, we use an \emph{adaptive} time increment according to the \emph{Courant-Friedrichs-Lewy} stability conditions for each of the physical processes (heat diffusion, metal percolation, radiation boundary condition) we consider in our work. Following \citet{Tok}, thermal conduction imposes the following critical time step:
\begin{equation}\label{eq:tstep}
\Delta t_{cond} = \frac{\left(\rho c\right)_m \Delta r^2}{2K_m}.
\end{equation}
In analogy with \citet{Tok}, we can define the following critical time step associated to the radiation boundary condition:
\begin{equation}\label{eq:trad}
\Delta t_{rad} = \frac{\left(\rho c\right)_m \Delta r T_{surf}}{\sigma\left( T^4_{surf} - T^4_0\right)}.
\end{equation}
Finally, while the percolation of metals is taking place, we need to solve also eq.(\ref{eq:chim}) and introduce a third critical time step:
\begin{equation}\label{eq:Courant}
\Delta t_{perc} = \frac{\Delta r}{v},
\end{equation}
where $v$ is the velocity of the metal percolation.
The Courant-Friedrichs-Lewy stability condition requires that the time step used in our model satisfies the following criterion:
\begin{equation}
\Delta t < min\left(\Delta t_{cond}, \Delta t_{rad}, \Delta t_{perc}\right).
\end{equation}
Therefore, at each temporal iteration of the program we select the actual time step based on the minimum critical time step among those we computed. As the stability condition requires the actual time step to be lower than the critical one, we chose to use a value equal to 90$\%$ of the smallest critical time step as a compromise between the competing needs for stability and performances. So, our time step is defined as:
\begin{equation}
\Delta t = 0.9 \times min\left(\Delta t_{cond}, \Delta t_{rad}, \Delta t_{perc}\right).
\end{equation}

\section{Results}

The results of our geophysical and thermal model depend on all the various physical parameters that we reported in Table \ref{tab:Parameters} and described previously. In the following discussion we will first focus on the dependence of the evolution of Vesta from the delay time ($\Delta t_d$). The delay time is essentially an unknown parameter, but it is critical in determining the initial overall abundance of short-lived radioactive elements, i.e. the intensity of the source of energy, the maximum temperature reached during the evolution and then the cooling behavior of the object. The chosen values of $\Delta t_d$ sample a long temporal interval and therefore very different intensities of the radioactive sources.
We investigated seven scenarios, labeled as N0-N6: as shown in Table \ref{tab:Scenario}, we considered values for $\Delta t_d$ equal to 0, 0.5, 1, 1.5, 2, 2.5 and 3 times the half-life of $^{26}$Al, corresponding to values in the range 0-2.15 Ma.

Before proceeding, we want to stress that the delay time ($\Delta t_d$) does not coincide with the accretion time of Vesta. The former represents the time between the injection of $^{26}$Al in the Solar Nebula (which in principle may not coincide with the time of the condensation of CAIs) and incorporates the latter, which however is always assumed negligible respect to $\Delta t_d$.

\begin{figure*}[H]
  \centering
  \includegraphics{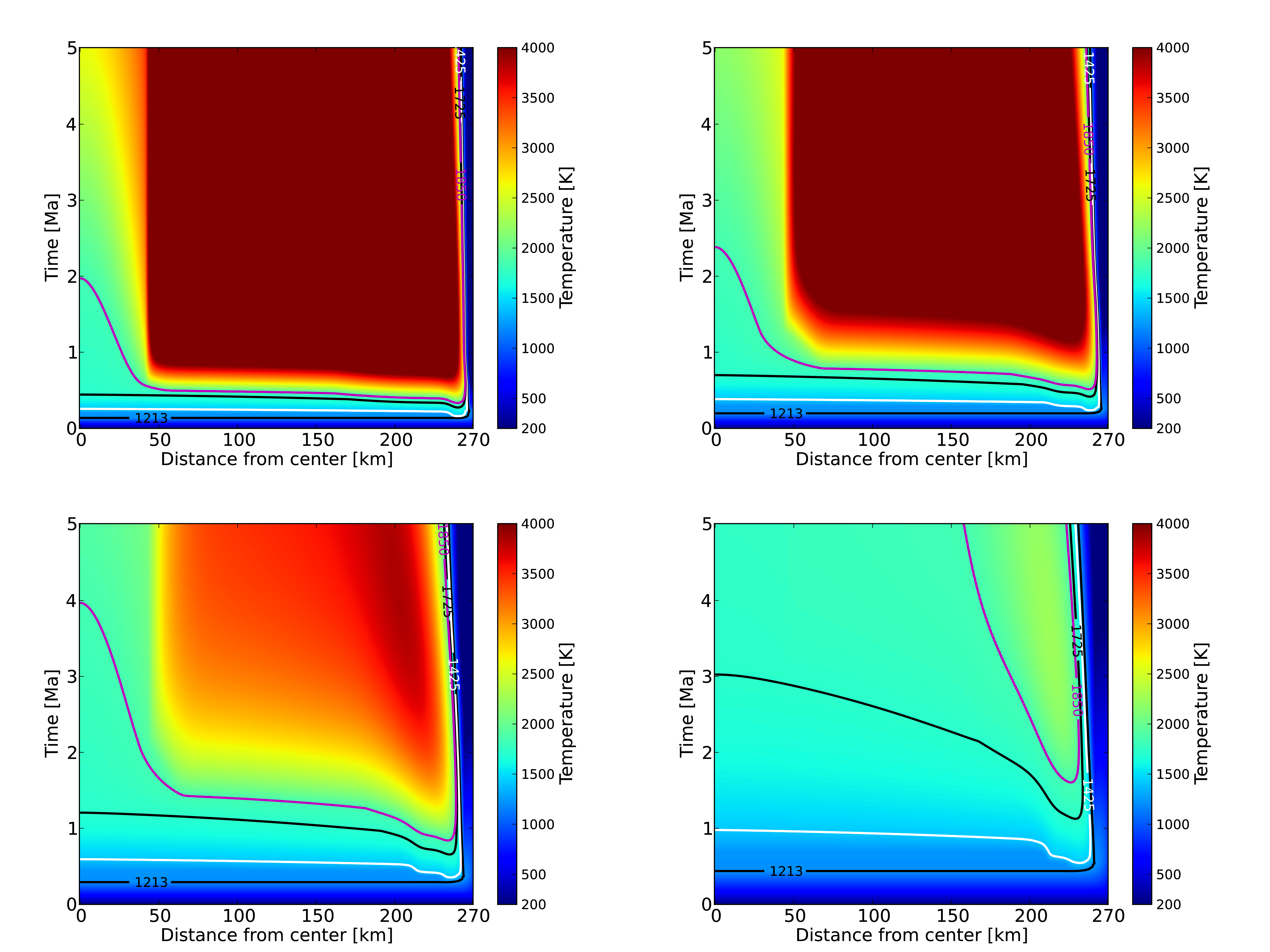}
  \caption{From left to right, from top to bottom: thermal history maps, for $\phi=1$ vol.$\%$, of N0 (a), N1 (b), N2 (c), N3 (d). Black isoline (1213 K) identifies the onset of metal melting; white (1425 K), black (1725 K) and magenta (1850 K) isolines identify the onset, the 50 vol.$\%$ and the complete melting of silicates, respectively. }\label{fig:HeatMap1}
\end{figure*}

\begin{figure*}[H]
  \centering
  \includegraphics{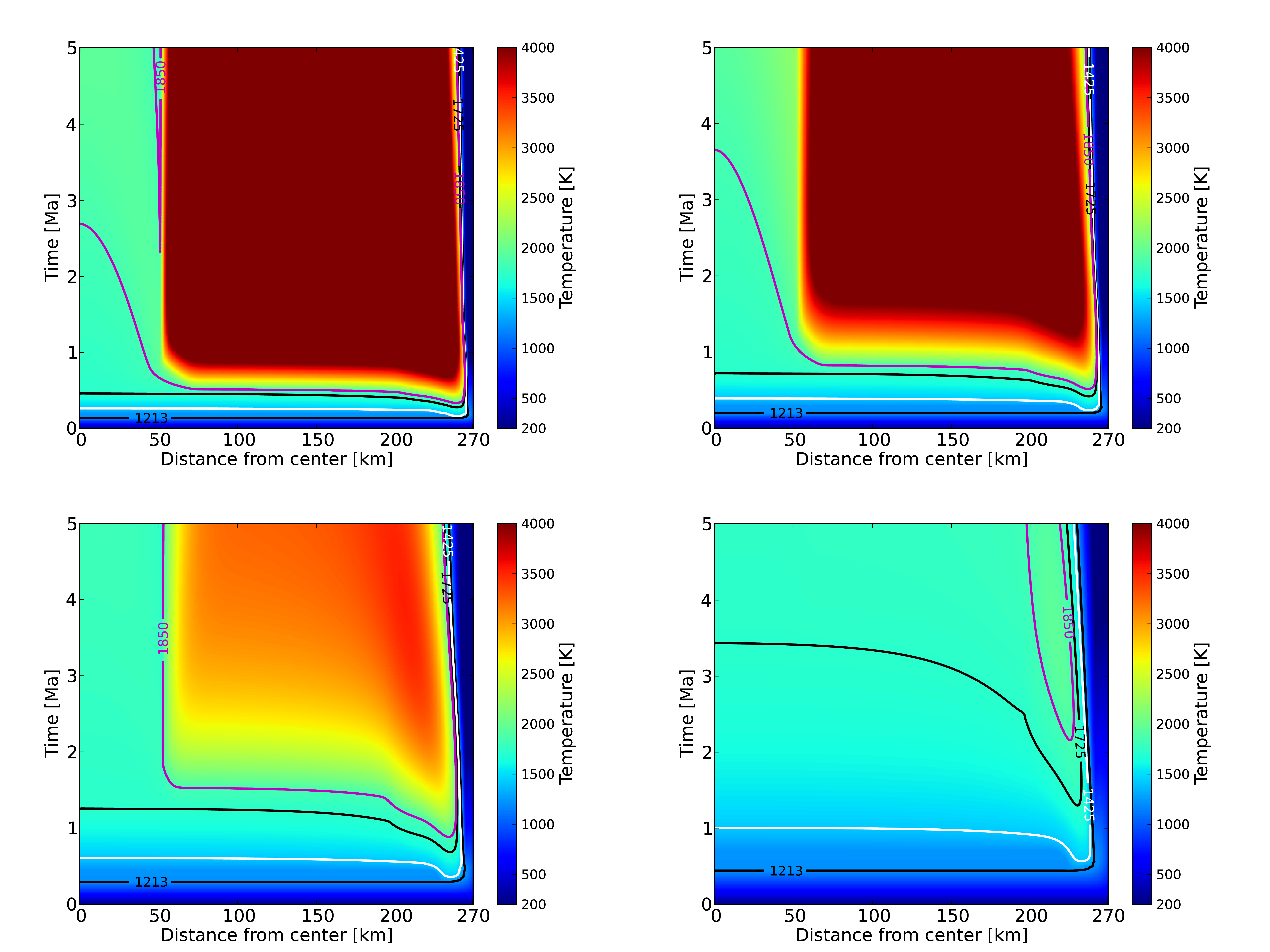}
  \caption{From left to right, from top to bottom: thermal history maps, for $\phi=2$ vol.$\%$, of N0 (a), N1 (b), N2 (c), N3 (d). Black isoline (1213 K) identifies the onset of metal melting; white (1425 K), black (1725 K) and magenta (1850 K) isolines identify the onset, the 50 vol.$\%$ and the complete melting of silicates, respectively. }\label{fig:HeatMap2}
\end{figure*}

\begin{figure*}[H]
  \centering
  \includegraphics{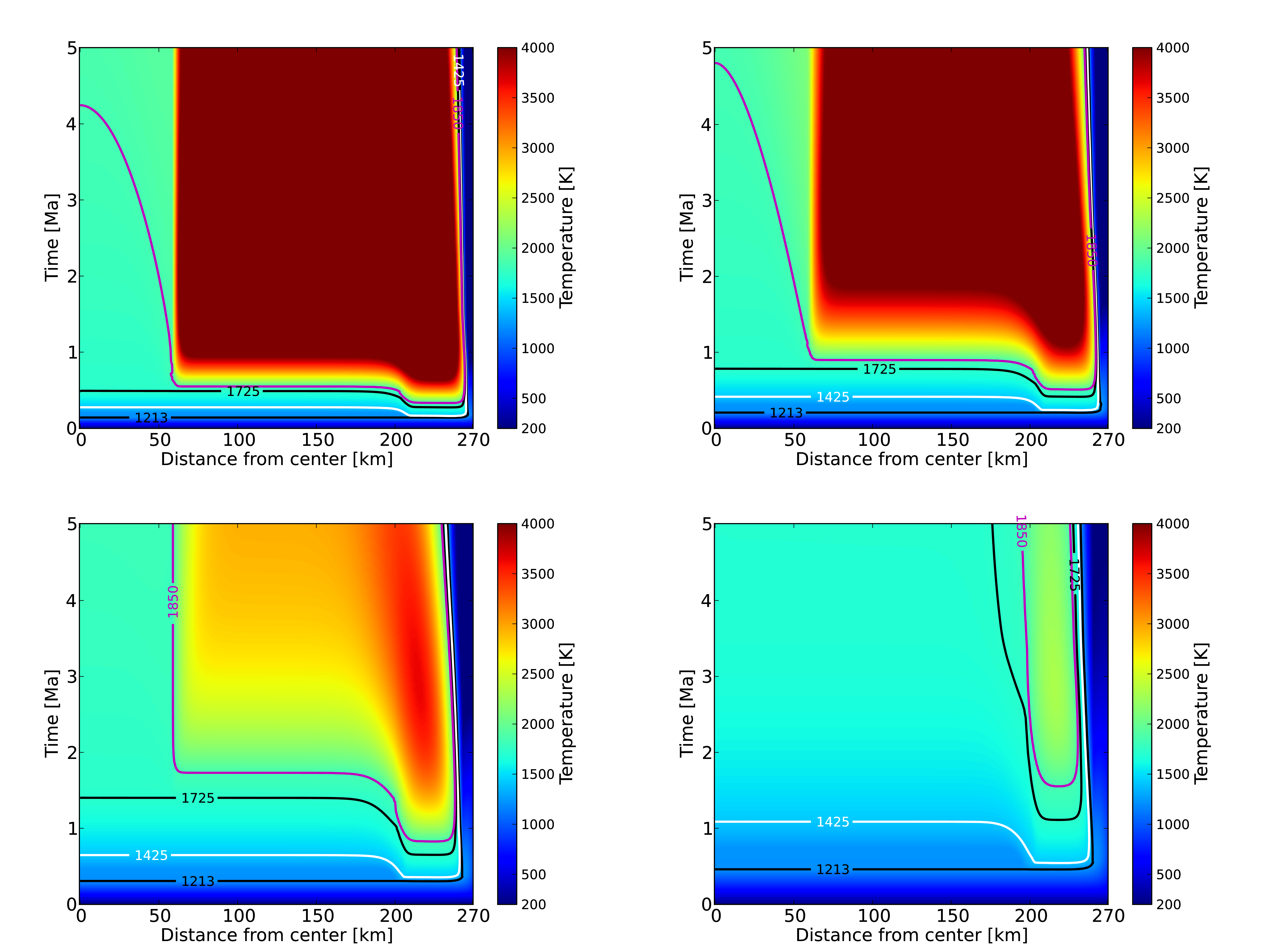}
  \caption{From left to right, from top to bottom: thermal history maps, for $\phi=5$ vol.$\%$, of N0 (a), N1 (b), N2 (c), N3 (d). Black isoline (1213 K) identifies the onset of metal melting; white (1425 K), black (1725 K) and magenta (1850 K) isolines identify the onset, the 50 vol.$\%$ and the complete melting of silicates, respectively.  }\label{fig:HeatMap5}
\end{figure*}

Our first analysis of the results obtained for the different scenarios was based on the compatibility of the simulated evolution of Vesta with the constrains supplied by the HEDs.
As we can see in Figs.\ref{fig:HeatMap1}, \ref{fig:HeatMap2} and \ref{fig:HeatMap5}, in N0-N2 scenarios, which imply delay times less than 1 Ma, the complete melting of silicates is achieved across the whole of the asteroid. In N3 scenario, where the delay time is 1 Ma, complete melting of silicates is achieved in a limited region of the mantle of Vesta but in the rest of the asteroid the degree of melting is larger than 50 vol.$\%$ with the only exception of the case where the porosity is equal to 5 vol.$\%$. Similarly, in N4 case where $\Delta t_d$ = 1.4 Ma, the degree of melting is generally lower than 50 vol.$\%$, except possibly in the limited region of the mantle. Finally, if $\Delta t_d >$ 1.5 Ma, silicate melting is either not possible or is limited only to a small region of the asteroid.

\begin{figure*}[H]
  \centering
  \includegraphics{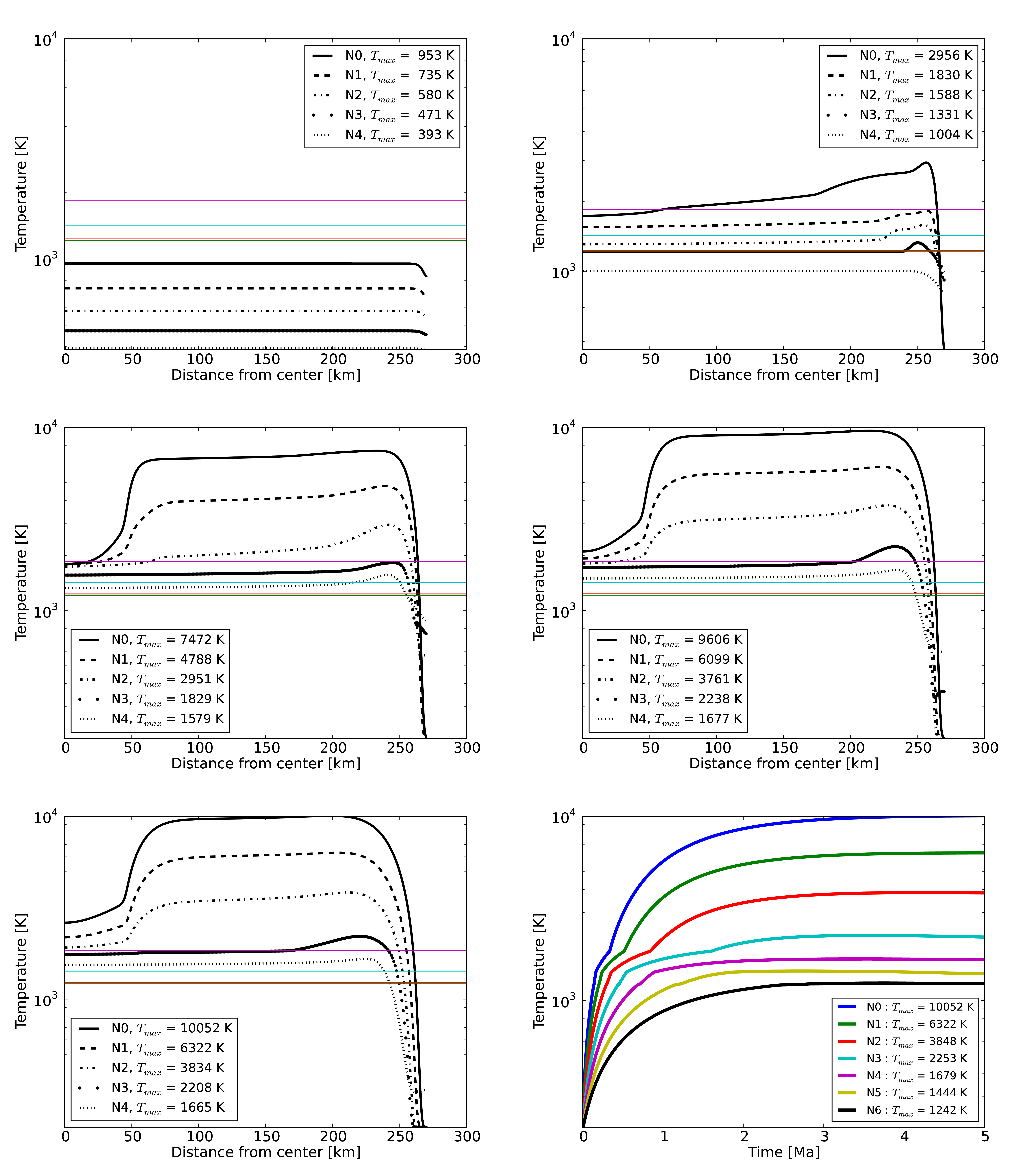}
  \caption{From left to right, from top to bottom, plots are: temperature vs distance from center at 0.1 Ma (a), at 0.5 Ma (b), at 1.5 Ma (c), at 3.0 Ma (d), at 5.0 Ma (e) and maximum temperature profile vs time (f), for $\phi=1$ vol.$\%$. The horizontal lines represent: the windows for the melting of metals (green and red) and the of the silicates (cyan and magenta).}\label{fig:Temp1}
\end{figure*}

\begin{figure*}[H]
  \centering
  \includegraphics{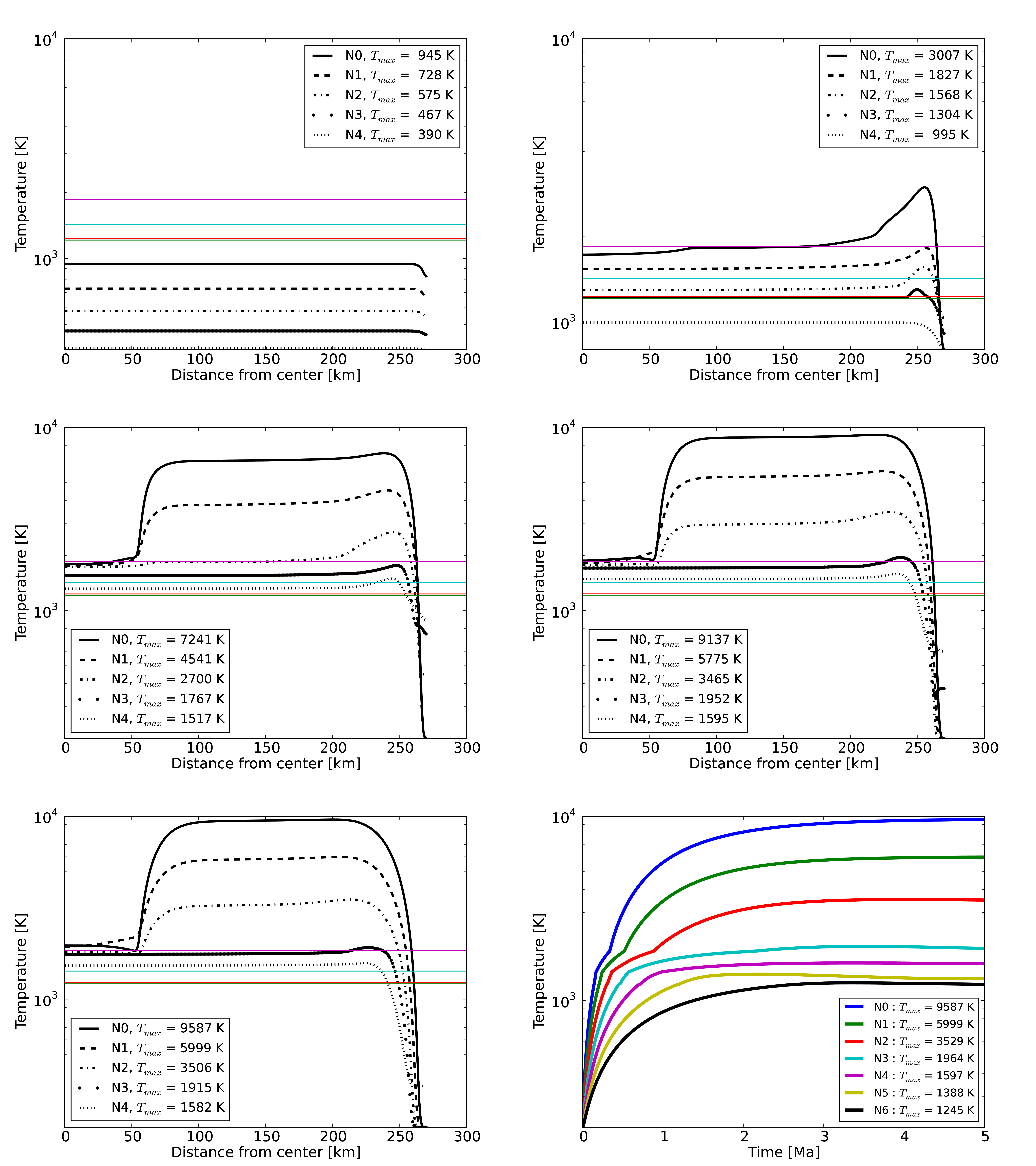}
  \caption{From left to right, from top to bottom, plots are: temperature vs distance from center at 0.1 Ma (a), at 0.5 Ma (b), at 1.5 Ma (c), at 3.0 Ma (d), at 5.0 Ma (e) and maximum temperature profile vs time (f), for $\phi=2$ vol.$\%$. The horizontal lines represent: the windows for the melting of metals (green and red) and the of the silicates (cyan and magenta).}\label{fig:Temp2}
\end{figure*}

\begin{figure*}[H]
  \centering
  \includegraphics{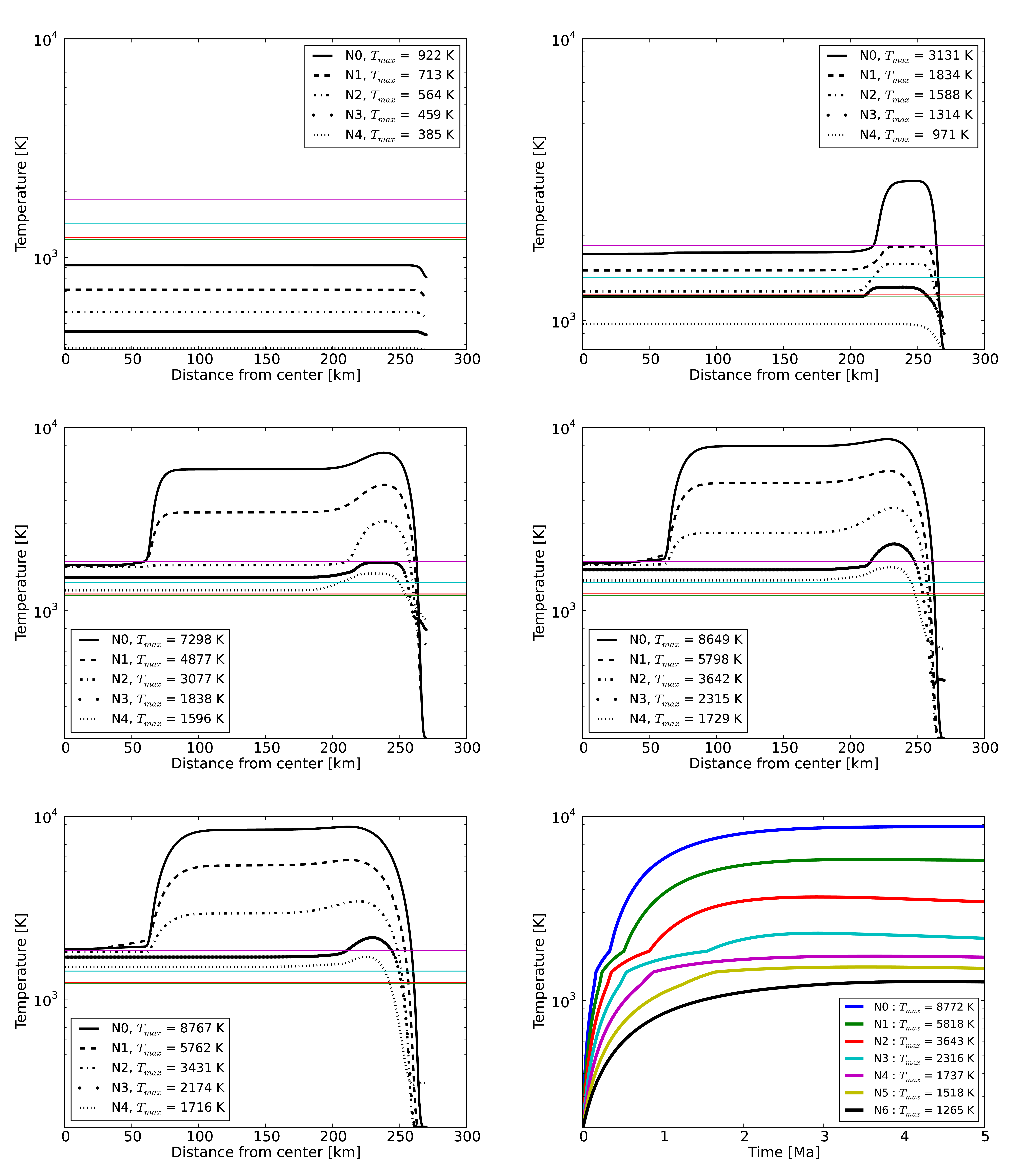}
  \caption{From left to right, from top to bottom, plots are: temperature vs distance from center at 0.1 Ma (a), at 0.5 Ma (b), at 1.5 Ma (c), at 3.0 Ma (d), at 5.0 Ma (e) and maximum temperature profile vs time (f), for $\phi=5$ vol.$\%$. The horizontal lines represent: the windows for the melting of metals (green and red) and the of the silicates (cyan and magenta).}\label{fig:Temp5}
\end{figure*}

The N0-N3 scenarios are compatible with the results of \citet{Green} which link the formation of eucrite and diogenite to large scale ($>50$ vol.$\%$) melting of the silicates. This would imply Vesta formed no later than 1 Ma after injection of $^{26}$Al. It is be noted that in those scenarios in which the differentiation takes place, the melting of the silicatic component begins in the first 1 Ma and the differentiation is completed in about 3 Ma. The formation of Vesta between 1 and 1.5 Ma after the injection of $^{26}$Al in the Solar Nebula is compatible with the formation of HEDs if eucrite and diogenites can form from a partial melt ranging to 25-50 vol.$\%$. Those cases where Vesta appeared in the Solar Nebula after more than 1.5 Ma since the injection of $^{26}$Al are not compatible with petrogenesis of HEDs. The conditions to start the formation of eucrites and diogenites are always obtained within 1 to 2 Ma from the formation of Vesta. In Figs.\ref{fig:HeatMap1}, \ref{fig:HeatMap2} and \ref{fig:HeatMap5} we set the maximum temperature to 4000 K to ease the comparison: we refer the reader to Figs.\ref{fig:Temp1}(e), \ref{fig:Temp2}(e) and \ref{fig:Temp5}(e) for the maximum temperature reached in the different cases.

We observe that the maximum temperature reached in the N0-N2 far exceeds the liquidus silicates melting temperature (1850 K): this happens because in our model we do not take into account other cooling mechanisms (convection and effusive phenomena) other than the conduction and the radiation at the surface. In Figs.\ref{fig:Temp1}, \ref{fig:Temp2}, \ref{fig:Temp5} for three different initial post-sintering porosities, we report in (a), (b), (c), (d) and (e) temperature profiles at different times (0.1, 0.5, 1.5, 3 and 5 Ma), in (f) the maximum temperature vs time profile. In (a), (b), (c), (d) and (e) the horizontal lines represent: the windows for the melting of metals (green and red) and silicates (cyan and magenta). Note that we report in Figs. \ref{fig:Temp1} - \ref{fig:Temp5} (in (a) - (e)) the profiles of N0-N4 scenarios while we neglect N5 and N6 because in these scenarios the differentiation of the asteroid does not take place. The general trend we observe in the different cases is the following. The first phase is characterized by a homogeneous heating of the asteroid. The second one is an increase of the temperature in the region in which the metal is depleting and migrating to the center of the asteroid. The final phase is the formation of the metallic core followed by the migration of the silicates (with the $^{26}$Al) towards the surface and the increase of the temperature in the mantle of the asteroid. When the differentiation occurs, the formation of a pure metallic core is possible. For a delay of about 1.5 Ma, the formation of the metallic core is possible only if the porosity is lower than 5 vol.$\%$.

We point out that, with the assumed chondritic composition and in the limits of 1D modeling, in those scenarios in which the differentiation takes place, the formation of a core of about 60 km (with a density of about 6200 kgm$^{-3}$) is possible. Its mass represents about the 2 vol.$\%$ of the total mass, slightly lower than the minimum value (4 vol.$\%$) given by \citet{Ruzicka}, while the moment of inertia ($MoI$) is 0.33.
The core size (60 km) is lower (about $1/2$) than Dawn's current estimates. As we can see in \citet{Russell}, assuming a core density of 7100 and 7800 kgm$^{-3}$, a core size of 107-113 km and a core mass fraction of about 18$\%$ were derived using a two-layer mass-balance model that can reproduce the gravitational moment $J_2$. The difference between Dawn's estimate of the core size and the current work is due to the 1-D geometry of our model. In such geometry all radial cells have the same unity volume, while in 3-D geometry the contribution of external shells is larger. Assuming an initial metal fraction of about 25 vol.$\%$, the maximum size of a pure metallic core is about 60 km. Probably, assuming a more iron-rich initial composition, the core would be larger and the moment of inertia would be different.

In Fig.\ref{fig:Temp1}(a) Vesta heats up maintaining an almost uniform temperature due to the initially homogeneous distribution of $^{26}$Al, in all the scenarios. Due to the strong heating source, after 0.5 Ma (see Fig.\ref{fig:Temp1}(b)), very high temperatures are reached in N0 and N1 and this trend continues after 1.5 Ma (see Fig.\ref{fig:Temp1}(c)) and also involves N2 scenario. In N3 and N4 silicate melting is still partial. After 3 Ma (see Fig.\ref{fig:Temp1}(d)), only in N4 the complete melting of silicates is not reached and the temperature profile is inside the melting temperature window of silicates. In Fig.\ref{fig:Temp1}(e), after 5 Ma, the general trend for all the profiles is similar to that after 3 Ma. We report in Fig.\ref{fig:Temp1}(f) the maximum temperature vs time and we observe that the differentiation is possible only in N0-N3 scenarios, for which the separation of silicates and metals occurs. The formation of the core takes place between 0.45 and 3.06 Ma (in N0-N3).

We can observe in Figs.\ref{fig:Temp2} and \ref{fig:Temp5} (in (a) - (e)) that the evolution of the temperature in the first 5 Ma from CAIs is quite similar to the previous case (i.e. porosity of 1 vol.$\%$). In the case of porosity of 2 vol.$\%$, after 5 Ma Vesta is in the cooling phase in N3 and N4, while if the porosity is equal to 5 vol.$\%$ the asteroid enters the cooling phase in all scenarios, except N0 and N1.

In our simulations we observed that the outermost region of Vesta never reaches the melting temperatures of either silicates or metals. This region would maintain the original composition of the pre-differentiated Vesta and we therefore named it the chondritic crust. Alongside the chondritic crust, we can derive from our simulations the thickness of the outer layer of material that is below the melting temperature of silicates (i.e. solid, SL in the following). Before the differentiation, the solid layer instantaneously coincides with the (shrinking) chondritic crust but, after differentiation occurs, it can be interpreted as the sum of the chondritic crust plus the solidifying eucritic and diogenitic layers. The evolution of the SL (chondritic crust) is identified, in the plots of Figs.\ref{fig:Temp1}, \ref{fig:Temp2} and \ref{fig:Temp5} by the outermost intersection between the temperature profile and the line corresponding to the onset of silicate (metal) melting. This behavior is present also in the modeling of \citet{GMC}, who however assumed this point to be fixed. In this work, instead, the location of this intersection moves in time depending on the characteristics of the scenarios (see Table \ref{tab:Results}).

In Table \ref{tab:Results}, we report the results obtained for different values of $\Delta t_d$ and post-sintering porosity, in particular the time at which the metal ($\tau_{met}$) melting begins (T = 1213 K), the time of formation of core ($\tau_{core}$), the time at which the silicate ($\tau_{sil}$) melting ends (T = 1850 K), the size of the surviving chondritic crust, the size of the SL after 3 and 5 Ma, the maximum degree of the silicate melting and the maximum temperature reached inside the asteroid after 5 Ma. The formation of the core (i.e. the separation of two melts) begins when the temperature reaches 1725 K in the region containing metals: $\tau_{core}$ indicates the time when all metals migrated toward the center.

In the case of 1 vol.$\%$ porosity, the thickness of the chondritic crust ranges from 3 to 17 km, while the thickness of the solid layer ranges from 7 to 21 km after 3 Ma and from 8 to 27 km after 5 Ma. In cases of porosity of 2 and 5 vol.$\%$, the chondritic crust ranges from 4 to 18 km and 4 to 19 km, respectively. After 3 Ma the SL ranges from 8 to 21 km and after 5 Ma the SL ranges from 9 to 30 km in both cases (see Table \ref{tab:Results}).

It is be noted that the temperature at the surface, while never reaching the solidus temperature of pure iron metal, can reach values as high as 1000 K. It is also noteworthy that the SL thickness implies a $\sim$5-10 km thick eucritic layer can already form between 3 and 5 Ma, in agreement with the dating of the oldest eucrites described by \citet{Bizzarro}.
\begin{table}[!h]
\centering{
\resizebox{0.4\textwidth }{!}{
\begin{tabular}[c]{|c|c|c|}
  \hline
  \textbf{Scenario}  & \textbf{Delay [Ma]} & \textbf{Delay[$\tau_{Al}$]}\\
  \hline
  N0 & 0.00 & 0.0\\
  N1 & 0.36 & 0.5\\
  N2 & 0.72 & 1.0\\
  N3 & 1.08 & 1.5\\
  \hline
  N4 & 1.43 & 2.0\\
  N5 & 1.79 & 2.5\\
  N6 & 2.15 & 3.0\\
  \hline
\end{tabular}}}
\caption{Seven scenarios have been studied, where the delay-parameter $\Delta t_d$ is expressed in Ma and in half-lives of $^{26}$Al.}\label{tab:Scenario}
\end{table}

\section{Conclusions}

In this paper we investigated the thermal history of Vesta across the first 5 Ma of its history. We focused our analysis on the effects of time passed between the injection of $^{26}$Al in the Solar Nebula and the formation of Vesta (which is assumed instantaneous), and consequently the content of $^{26}$Al, and of the initial porosity on the subsequent evolution of the asteroid. Our model is based on the simultaneous study of the evolution of the internal structure (percolation of metals and separation of silicatic and metallic melts) and the thermal heating of the asteroid, due to $^{26}$Al and other short and long-lived radionuclides. We observed that the main source of energy is represented by $^{26}$Al while the contribution of the other radionuclides is negligible.

At the surface, the balance between thermal heating and the black-body radiation into space offers a more realistic picture of the thermal history than a simple case in which the temperature of the surface is fixed to a constant value. We explored several scenarios (N0-N6) characterized by different radiogenic strengths, expressed by a delay-parameter ($\Delta t_d$) ranging from 0 to 2.16 Ma. We opted for an initial composition of about 75 vol.$\%$ of silicates and about 25 vol.$\%$ (similar to H or L chondrite compositions): the chosen values of the porosity are 1, 2 and 5 vol.$\%$. In the scenarios in which the differentiation takes place a metallic core is always formed, its formation time ranging from about 0.5 to 3.5 Ma. We also observed the solidification of a surface layer whose maximum thickness is about 30 km. This solid layer is composed by an unmelted chondritic crust with its thickness reaching the maximum of about 20 km and the exact value strongly depending on the delay time ($\Delta t_d$). A $\sim$5-10 km thick layer (which can be interpreted as the topmost part of the eucritic layer observed by Dawn) can already form beneath the chondritic crust between 3 and 5 Ma, in agreement with the dating of the oldest eucrites described by \citet{Bizzarro}. The solidification of this layer is possible even in absence of convection and volcanism, but only considering conduction of heat and radiation at the surface.

It is interesting to note that the existence of a chondritic crust overlying a differentiated interior was speculated by \citet{Elkins11}, based on analytic arguments, for the other two largest bodies in the asteroid belt, Ceres and Pallas, but not for Vesta. Moreover, the independent results obtained by \citet{TGB} using a 2D finite difference code confirm our findings: while not explicitly discussed in their paper, their Figs. 4 and S2 clearly show the existence of a surface layer of material that never experiences melting (i.e. our chondritic crust) whose thickness varies between 3 and 7 km. A further comparison between the two models also reveals an agreement between the constrains on the formation time of Vesta (G. Golabek, personal communication), i.e. Vesta's formation should have occurred preferably before 1 Ma and, more generally, no later than 1.5 Ma respect to the injection of $^{26}$Al in the Solar Nebula (which, in their work, coincides with CAIs crystallization).

The survival of the chondritic crust and the rate of solidification of the underlying eucritic layer are particularly important to understand the geophysical and thermal evolution of Vesta as, across the temporal interval here investigated, the asteroid underwent a bombardment caused by the formation of Jupiter (\Citealt{Turrini1,Turrini2}). The Jovian Early Bombardment, in fact, causes a global erosion of the primordial crust of Vesta (\Citealt{Turrini3}) and the chondritic crust, not currently observed by Dawn, plays an important role in preserving the eucritic layer. Moreover, the Jovian Early Bombardment can trigger local or large-scale effusive phenomena due to the excavation of craters and the formation of impact basins (\Citealt{Turrini1,Turrini3}), thus affecting the cooling history of Vesta.

The complete melting of silicates is achieved if Vesta does not form later than 0.8 Ma after the injection of $^{26}$Al in the Solar Nebula while it is achieved in a limited region of the mantle if the formation took place after 1 Ma: in the rest of the asteroid a 50 vol.$\%$ melting of silicates occurs. If the formation is completed no later than 1.4 Ma from $^{26}$Al injection the degree of melting is generally lower than 50 vol.$\%$. For delay times of more than 1.5 Ma the melting, when occurring, is limited in a small region of the asteroid.

The formation of Vesta should not have occurred later than 1 Ma after $^{26}$Al injection, if the crystallization of eucrites and diogenites are linked to a large degree (more than 50 vol.$\%$) of silicate melting (\Citealt{Green}). If eucrites and diogenites can form from a partial melt ranging to 25-50 vol.$\%$ the formation of Vesta could have occurred up to 1.4 Ma after $^{26}$Al injection.

\section*{Acknowledgements}

We wish to thank Guy J. Consolmagno, Sin-Iti Sirono and Gal Sarid for their helpful suggestions and comments and Romolo Politi for his numerical analysis assistance.
We also wish to thank Harry Y. McSween, Sandeep Sahijpal, Gregor Golabek, Paul Tackley and two anonymous referees for their valuable comments.
The computational resources used in this research have been supplied by INAF-IAPS through the project HPP-High Performance Planetology. This work has been supported by the International Space Science Institute in Bern through the International Teams project ``Vesta, the key to the origins of the Solar System".

This work is dedicated to \emph{Angioletta Coradini}.

\begin{table*}[H]
\centering
\resizebox{1\textwidth }{!}{
\begin{tabular}[c]{|c|c|c|c|c|c|c|c|c|}
  \hline
   & \textbf{$\tau_{met}$} \textbf{[Ma]} & \textbf{$\tau_{core}$} \textbf{[Ma]} & \textbf{$\tau_{sil}$} \textbf{[Ma]}&  \textbf{Chondritic crust} \textbf{[km]} & \textbf{$SL_{3Ma}$} [\textbf{km}] & \textbf{$SL_{5Ma}$} [\textbf{km}] &\textbf{vol.$\%$ Silicate melt} & \textbf{$T_{max}$} \textbf{[K]}\\
  \hline
  \textbf{N0} &&&&&&&&\\
  $\phi = 1.0$ vol.$\%$& 0.14 & 0.45 & 0.34 & 3 & 7 & 8 & 100 & $\simeq10000$\\
  $\phi = 2.0$ vol.$\%$& 0.14 & 0.47 & 0.34 & 4 & 8 & 9 & 100 & $\simeq9500$\\
  $\phi = 5.0$ vol.$\%$& 0.15 & 0.51 & 0.43 & 4 & 8 & 9 & 100 & $\simeq8700$\\
  \hline
  \textbf{N1} &&&&&&&&\\
  $\phi = 1.0$ vol.$\%$& 0.21 & 0.72 & 0.52 & 5 & 10 & 11 & 100 & $\simeq6500$\\
  $\phi = 2.0$ vol.$\%$& 0.21 & 0.73 & 0.52 & 5 & 10 & 13 & 100 & $\simeq6000$\\
  $\phi = 5.0$ vol.$\%$& 0.22 & 0.80 & 0.52 & 6 & 11 & 13 & 100 & $\simeq 5800$\\
  \hline
  \textbf{N2} &&&&&&&&\\
  $\phi = 1.0$ vol.$\%$& 0.30 & 1.20 & 0.84 & 7 & 12 & 17 & 100 & 3848\\
  $\phi = 2.0$ vol.$\%$& 0.31 & 1.27 & 0.88 & 7 & 12 & 17 & 100 & 3529\\
  $\phi = 5.0$ vol.$\%$& 0.31 & 1.43 & 0.83 & 8 & 13 & 18 & 100 & 3643\\
  \hline
  \textbf{N3} &&&&&&&&\\
  $\phi = 1.0$ vol.$\%$& 0.46 & 3.06 & 1.60 & 11 & 15 & 20 & 100 & 2253\\
  $\phi = 2.0$ vol.$\%$& 0.47 & 3.42 & 2.20 & 11 & 16 & 20 & 100 & 1964\\
  $\phi = 5.0$ vol.$\%$& 0.47 & - -  & 1.56 & 11 & 16 & 20 & 100 & 2316\\
  \hline
  \textbf{N4} &&&&&&&&\\
  $\phi = 1.0$ vol.$\%$& 0.72 & - - & - - & 17 & 21 & 27 & $\simeq40$ & 1679\\
  $\phi = 2.0$ vol.$\%$& 0.73 & - - & - - & 18 & 21 & 30 & $\simeq35$ & 1597\\
  $\phi = 5.0$ vol.$\%$& 0.74 & - - & - - & 19 & 21 & 30 & $\simeq50$ & 1737\\
  \hline
\end{tabular}}
\caption{Summary of scenarios. We report the time at which the metal melting occurs ($\tau_{met}$, T = 1213 K), the time of formation of the core ($\tau_{core}$), the time at which the silicate melting ends ($\tau_{sil}$, T = 1850 K), the chondritic crust thickness, the solid layer thickness after 3 and 5 Ma, the percentage of silicate melt and the maximum temperature reached after 5 Ma. }\label{tab:Results}
\end{table*}

\end{document}